\documentclass[pra,notitlepage,twocolumn]{revtex4}

\usepackage{amsmath}
\usepackage{amssymb}
\usepackage{bm}
\usepackage{graphicx}
\usepackage{times}

\begin{document}

\title{Optical properties of honeycomb photonic structures}

\author{Artem~D.~Sinelnik${}^{1}$}
\author{Mikhail~V.~Rybin${}^{2,3}$}
\email{m.rybin@mail.ioffe.ru}
\author{Stanislav~Y.~Lukashenko${}^{1}$}
\author{Mikhail~F.~Limonov${}^{2,3}$}
\author{Kirill~B.~Samusev${}^{2,3}$}

\affiliation{$^1$Department of Nanophotonics and Metamaterials, ITMO University, St.~Petersburg 197101, Russia}
\affiliation{$^2$Ioffe Institute, St.~Petersburg 194021, Russia}
\affiliation{$^3$Department of Dielectric and Semiconductor Photonics, ITMO University, St.Petersburg 197101, Russia}

\begin{abstract}
We study, theoretically and experimentally, optical properties of different types of honeycomb photonic structures, known also as `photonic graphene'. First, we employ the two-photon polymerization method to fabricate the honeycomb structures. In experiment, we observe a strong diffraction from a finite number of elements, thus providing a unique tool to define the exact number of scattering elements in the structure by a naked eye. Then, we study theoretically the transmission spectra of both honeycomb single layer and 2D structures of parallel dielectric circular rods. When the dielectric constant of the rod materials $\varepsilon$ is increasing, we reveal that a two-dimensional photonic graphene structure transforms into a metamaterial when the lowest TE${}_{01}$ Mie gap opens up below the lowest Bragg bandgap. 
%This transformation allows the homogenization of the two-dimensional honeycomb structure with the negative effective permeability at higher values of $\varepsilon$.
We also observe two Dirac points in the band structure of 2D photonic graphene at the $K$ point of the Brillouin zone and demonstrate a manifestation of the Dirac lensing for the TM polarization. The performance of the Dirac lens is that the 2D photonic graphene layer converts a wave from point source into a beam with flat phase surfaces at the Dirac frequency for the TM polarization.
\end{abstract}

\date{\today}% It is always \today, today,
             %  but any date may be explicitly specified

\maketitle

%==========================================================================
\section{Introduction}

The study of honeycomb structures being known for thousands of years is increased in the last few years due to the unique physical properties of graphene \cite{neto2009electronic}. Graphene, a two-dimensional honeycomb lattice of carbon atoms, has attracted enormous attention since its first discovery by isolation from bulk graphite using adhesive tape. Graphene and graphene-based materials demonstrated a very fast development of both fundamental and practical aspects in optics and electronics \cite{chang2013graphene}. A new research hotspot becomes a Dirac point that was first investigated in the electronic energy band structure of graphene \cite{novoselov2005two}. Importantly, this concept penetrated into the field of optics where the so-called photonic graphene, a two-dimensional photonic crystal structure that is analogous to graphene, has been studied theoretically and extensively \cite{peleg2007conical,sepkhanov2007extremal,polini2013artificial,rechtsman2013topological,plotnik2014observation}

The conical diffraction and the dynamics of optical waves in photonic graphene was studied experimentally \cite{peleg2007conical,diebel2016conical}. In particular, it was demonstrated that an incident narrow light beam at 488 nm wavelength, with momentum at the vicinity of a diabolic point, diffracts in the honeycomb lattice in a characteristic conical form, obtaining the shape of a ring whose thickness does not broaden, whereas its radius grows linearly with distance \cite{peleg2007conical}. An unconventional edge state at the zigzag edge of an optically induced honeycomb lattice residing on the bearded edge was demonstrated experimentally \cite{plotnik2014observation}. Very recently, a novel kind of dispersive edge state was found in a photonic honeycomb lattice experimentally that emerges not only in zigzag and bearded terminations, but also in armchair edges \cite{milicevic2017orbital}. At a microwave frequency, the measured transmission through a structure of photonic graphene enters a pseudodiffusive regime where the transmission scales inversely with the thickness of the crystal \cite{zandbergen2010experimental}. The observation of a pseudodiffusive transport regime in a photonic graphene is one of the physical phenomena that can be observed in the photonic as well as in the electronic realization of graphene.

Photonic properties of a 2D photonic crystal with honeycomb lattice fabricated in silicon using electron-beam lithography were calculated numerically and measured experimentally \cite{ye2004silicon}. Lattice parameters of the fabricated structures made it possible to achieve and confirm experimentally the complete photonic band gap in vicinity of the optical communications wavelengths around 1.5 $\mu$m. Note that the photonic band structure of a 2D triangular lattice of air rods in a dielectric medium with high $\varepsilon $ can be found in textbooks \cite{g101}. For the particular radius $r/a = 0.48$ and dielectric constant $\varepsilon = 13$, TM and TE band gaps overlap, resulting in an 18.6\% complete photonic band gap. Also the effect of a Dirac point on the transmission spectra was calculated for a 2D photonic crystal of air rods formed a triangular lattice \cite{sepkhanov2007extremal}. Recently it is reported that a photonic crystal with honeycomb-like lattice supports a Dirac mode with a different algebraic-decay field profile \cite{mao2017light}.

Nevertheless, the optical studies of photonic graphene are a long way still from completeness. Neither the photonic band structure and transmission spectra of 2D crystal with honeycomb lattice at high dielectric constant $\varepsilon $ nor optical diffraction from photonic graphene were studied theoretically or experimentally. To improve the situation, in this paper, we calculate the transmission spectra of photonic graphene, investigate optical Laue diffraction both theoretically and experimentally and, finally, perform a comprehensive studies of photonic properties of 2D honeycomb lattice of parallel dielectric circular rods.

\section{Fabrication of photonics lattices}

%%%%%%%%%%%%%%%%%%%%%%%%%%%%
%% Figure 1
\begin{figure*}[!t]
\includegraphics[width=12cm]{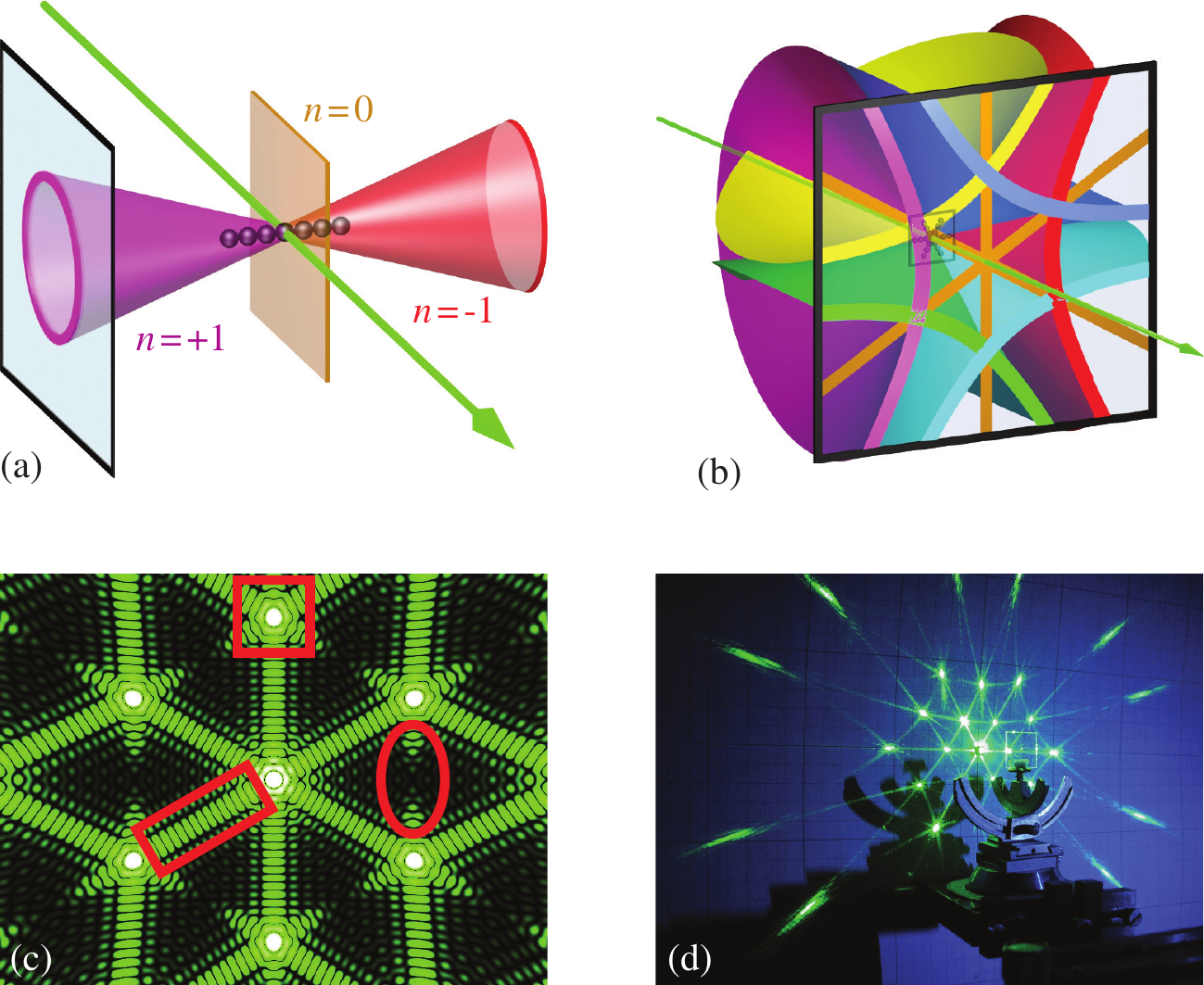}
\caption{ (a) Schematic of the zero-order ($n=0$) and first-order ($n=\pm 1$) Laue diffraction from a horizontally oriented chain of scatterers. (b) Schematic of the zero- and first-order Laue diffraction from a hexagonal structure. (c) Fragment of the diffraction pattern of honeycomb structure with its characteristic features discussed in text. (d) Photograph of experimental setup.}
\label{fig:schemes}
\end{figure*}
%
%%%%%%%%%%%%%%%%%%%%%%%%%%%%

Synthesis of graphene-based nanomaterials includes mechanical exfoliation from graphite, epitaxial growth from thermal deposition of SiC, chemical vapor deposition, solution-based exfoliation from graphite oxide and graphite, bottom-up organic synthesis methods \cite{chang2013graphene}. In this study, to fabricate photonic counterpart of graphene, a honeycomb structure, we employ two-photon polymerization method which is also called direct laser writing \cite{kawata2001finer,farsari2009materials}. This method is based on the nonlinear two-photon polymerization of a photosensitive material in the focus of a femtosecond laser beam and makes it possible to form a photonic structure with a transverse resolution below 100 nm \cite{li2009achieving}. In addition, we fabricate 2D photonic structures of hexagonal symmetry (constituted by dense-packed triangles). As this was done in our preceding studies, we use the installation and software package from Laser Zentrum Hannover (Germany) \cite{rybin2015band}. The structures are fabricated from a hybrid organic-inorganic material based on zirconium propoxide and an Irgacure 369 photo-initiator (Ciba Specialty Chemicals Inc., Switzerland). The polymerization is performed with a train of femtosecond pulses (wavelength was 780 nm) at a repetition frequency of 80 MHz (12.5 ns between consecutive pulses) from a 50-fs TiF-100F laser (Avesta-Project, Russia). The sample is mounted on a two-coordinate motorized linear air-bearing translator (Aerotech Inc., USA), by which the laser focus is moved along a 2D scanning path in the XY-plane. The laser radiation is focused in the photoresist volume through the glass substrate with a 100x oil-immersion microscope objective with numerical aperture NA = 1.4 (Carl Zeiss MicroImaging GmbH, Germany). The correspondence of the resulting materials to the designed structures is confirmed by scanning electron microscopy (SEM). In the structures, the lattice parameters are within the range $0.5\mu$m $\leqslant a \leqslant 2.0\mu$m in different samples and the number of scatterers varied from tens to tens of thousands.

\section{Laue diffraction from honeycomb structure}

The samples produced by the two-photon polymerization method are characterized by a low dielectric permittivity contrast. In this case, the optical diffraction is described in the Born approximation \cite{g341} when the diffraction intensity is determined by a product of the squared structural factor $S(\mathbf{q})$, scattering form factor $F(\mathbf{q})$ and polarization factor \cite{rybin2013dimensionality}. Our calculation in which only the structural factor is taken into account allows us to adequately describe the experimental data, and, therefore, we disregard the contributions from the form factor and the polarization factor. In the 2D case, the diffraction is determined by $S(\mathbf{q})=\sum _{i} \exp (i\mathbf{q}\mathbf{r}_{i} )$, where $\mathbf{r}_{i} =\mathbf{a}_{1} n_{1} +\mathbf{a}_{2} n_{2} $, $\mathbf{q}\equiv \mathbf{k}_{i} -\mathbf{k}_{s} $ is the scattering vector, $\mathbf{k}_{i} $ and $\mathbf{k}_{s} $ are the wave vectors of the incident and scattering waves. The structural factor has maxima in the scattering directions $\mathbf{k}_{s} $ determined by the system of Laue equations: $(\mathbf{k}_{s} -\mathbf{k}_{i} )\cdot \mathbf{a}_{1,2} =2\pi n_{1,2} $. For a 1D linear chain of point-like scatterers $\mathbf{r}_{n} =n\mathbf{a}_{1} $, one has $|S(\mathbf{q})|^{2} =\sin ^{2} (N\mathbf{q}\mathbf{a}/2)/(\mathbf{q}\mathbf{a}/2)$ \cite{guinier1963xray}. When the denominator becomes zero, $\sin (\mathbf{q}\mathbf{a}/2)\to 0$, the structural factor $|S(\mathbf{q})|^{2} $ has strong maxima named ``principal'' maxima in what follows. Figure~\ref{fig:schemes}c depicts the diffraction pattern of a calculated honeycomb structure, which shows seven principal diffraction peaks, one of which is marked by a red square. When the numerator goes to zero, $\sin^{2} (N\mathbf{q}\mathbf{a}/2)=0$, the function $|S(\mathbf{q})|^{2} $ becomes zero, and this means that it will have $N-1$ zeroes in the interval between the principal maxima and, accordingly, $N-2$ more maxima, named ``additional'' in what follows. These additional maxima are marked in Fig.~\ref{fig:schemes}c by a red rectangle. It is easy to obtain the conditions for the appearance of the principal maxima in the case of 1D chain and normal incidence:
\begin{equation} \label{eq:thetatheta}
\theta _{s} =\cos^{-1} \left(n\frac{\lambda }{a} \right),
\end{equation}
where $\lambda $ is the wavelength of incident light, and $\theta _{s} $ is the angle of light scattering on the chain.

\begin{figure*}[!t]
\includegraphics[width=12cm]{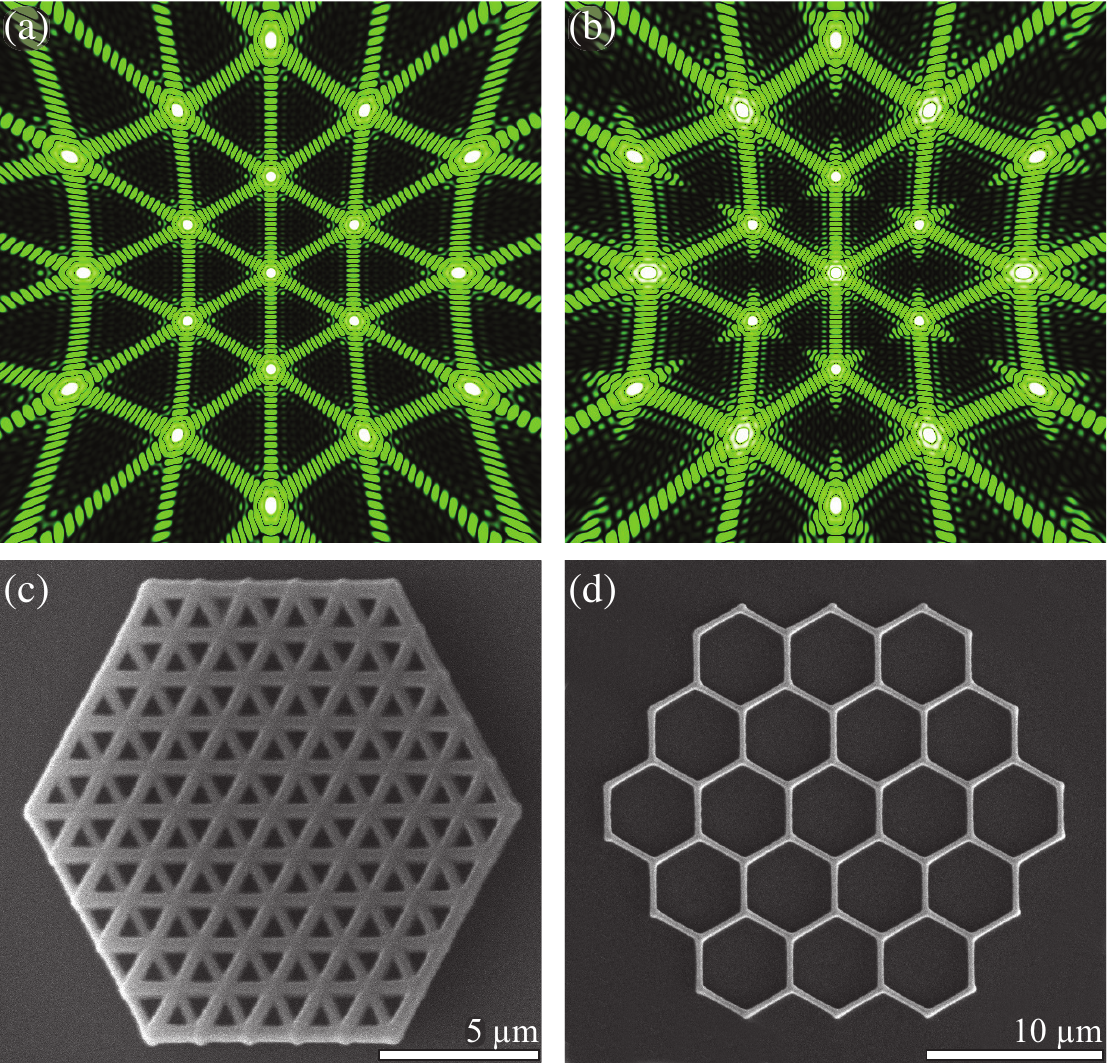}
\caption{Calculated diffraction patterns for triangular (a) and honeycomb (b) photonic structures on a flat screen positioned behind the sample. SEM images of the triangular (c) and honeycomb (d) photonic structures fabricated by the two-photon polymerization method.}
\label{fig:SEM}
\end{figure*}

Formula (\ref{eq:thetatheta}) sets the diffraction conditions in relation to the ratio between $\lambda $ and $a$ because the inverse cosine function is only defined in the interval from -1 to 1. The zero order of scattering ($n=0$) is always observed, with $\theta _{s} =90^\circ$, i.e., the diffraction always occurs in the plane perpendicular to the chain axis [Fig.~\ref{fig:schemes}a]. The next diffraction orders ($n=\pm 1$, $\pm 2$, $\ldots $) appear when the relation $|n|\lambda \le a$ is valid and have the form of pairs of cones having axis coinciding with the chain axis $a$ and apex angle $\theta _{s} $. In the experiments, we use a Nd-laser with $\lambda =0.53\mu $m, and, therefore, only the zero diffraction order be observed experimentally at $a<0.53\mu$m, whereas at $0.53\mu$m $<a<1.06\mu$m, a pair of first-diffraction order cones will be observed (Fig.~\ref{fig:SEM}a), and so on.

%%%%%%%%%%%%%%%%%%%%%%%%%%%%
%% Figure 2
%
%%%%%%%%%%%%%%%%%%%%%%%%%%%%

%%%%%%%%%%%%%%%%%%%%%%%%%%%%
%% Figure 3
\begin{figure}[!t]
\includegraphics{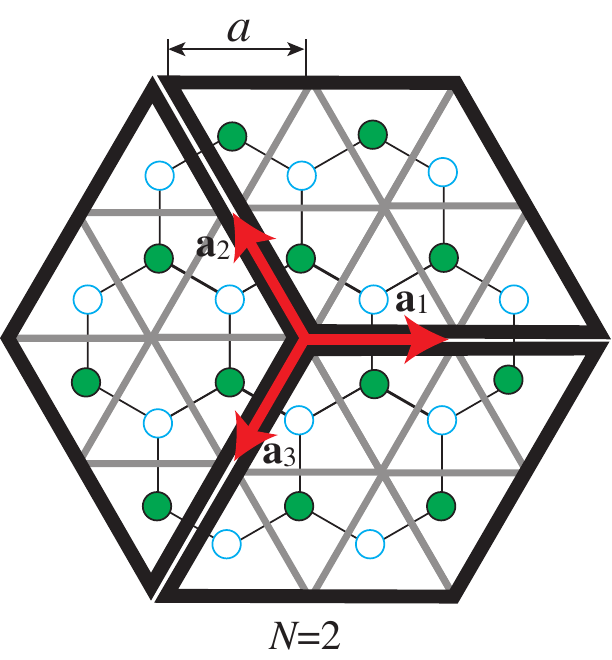}
\caption{ Schematic of the honeycomb photonic structure showing a subdivision of the hexagon into three parallelograms. The hexagon side is equal in length to two triangle sides $Na=2a$. $\mathbf{a}_{i} $ are the basis vectors of the hexagonal lattice. White and green circles correspond to the different triangular sub-lattices.}
\label{fig:GraphDiv}
\end{figure}
%
%%%%%%%%%%%%%%%%%%%%%%%%%%%%

Now we discuss a 2D photonic structure with the hexagonal symmetry C$_{6v} $ and one scatterer per unit cell, as in the case of a triangular tiling of the plane. Although the lattice $\mathbf{r}_{i} =\mathbf{a}_{1} n_{1} +\mathbf{a}_{2} n_{2} $ can be represented as a superposition of two systems of linear chains $\mathbf{r}_{1} =\mathbf{a}_{1} n_{1} $ and $\mathbf{r}_{2} =\mathbf{a}_{2} n_{2} $, for a triangular lattice it is convenient to consider three basis vectors $\mathbf{a}_{1} $, $\mathbf{a}_{2} $, and $\mathbf{a}_{3} $ because all the three directions in the hexagonal lattice are equivalent. To calculate the structural factor $S(\mathbf{q})$, we subdivide the hexagon into three parallelograms formed by three pairs of vectors ($\mathbf{a}_{1} $, $\mathbf{a}_{2} $), ($\mathbf{a}_{2} $, $\mathbf{a}_{3} $), and ($\mathbf{a}_{3} $, $\mathbf{a}_{1} $) (Fig.~\ref{fig:GraphDiv}). In our calculations, a hexagon is defined by the lattice constant $a$ and by the number $N$ of triangles it is formed from. The structural factor $S_{12} (\mathbf{q})$ for ($\mathbf{a}_{1} $, $\mathbf{a}_{2} $)-parallelogram can be written as:
\begin{equation}
\label{eq:Sfactor12}
S_{12} (\mathbf{q})=\frac{\sin (N\mathbf{q}\mathbf{a}_{1} /2)}{\sin (\mathbf{q}\mathbf{a}_{1} /2)} \frac{\sin (N\mathbf{q}\mathbf{a}_{2} /2)}{\sin (\mathbf{q}\mathbf{a}_{2} /2)} e^{-i\frac{(N-1)\mathbf{q}\mathbf{a}_{3} }{2}},
\end{equation}
and $S_{23} (\mathbf{q})$ and $S_{31} (\mathbf{q})$ can be found by the cyclic interchange of vectors $\mathbf{a}_{1} $, $\mathbf{a}_{2} $, and $\mathbf{a}_{3} $. For the whole triangular lattice, we combine $S_{ij} (\mathbf{q})$ by taking into account the origin of coordinates for all parallelograms and, as a result, obtain:
\begin{equation}
\label{eq:Sfactor}
S(\mathbf{q})=e^{i\frac{\mathbf{q}\mathbf{a}_{1} }{2} }S_{12} (\mathbf{q})+e^{-i\frac{\mathbf{q}\mathbf{a}_{1} }{2} }S_{23} (\mathbf{q})+e^{i\frac{\mathbf{q}(\mathbf{a}_{3} -\mathbf{a}_{2} )}{2}}S_{31} (\mathbf{q}).
\end{equation}

On the assumption that the $x$-component of $\mathbf{k}_{s} $ is zero, it is easy to demonstrated that the squared modulus of the structural factor along a vertical plane normal to the screen is following
\begin{equation}
\label{eq:SfHex}
\left| S(\mathbf{q})\right|^{2} \approx N^{2} \frac{\sin^{2} (2N\zeta )}{\sin ^{2} (\zeta )}
,
\end{equation}
where $\zeta =\left|\mathbf{k}_{s} \right|\sin \varphi \sqrt{3} a/4$, and $\varphi $ is the angle between $\mathbf{k}_{s} $ and $k_{i} $.

The triangular lattice can be represented as a superposition of three systems of linear chains, $\mathbf{r}_{1} =\mathbf{a}_{1} n_{1} $, $\mathbf{r}_{2} =\mathbf{a}_{2} n_{2} $, and $\mathbf{r}_{3} =\mathbf{a}_{3} n_{3} $, with the vectors $\mathbf{a}_{1} $, $\mathbf{a}_{2} $, and $\mathbf{a}_{3} $ turned relative to each other by an angle of 120$^{\circ } $. Thus, based on the above analysis of the diffraction on a 1D linear chain of scatterers, we would expect for the triangular lattice the appearance of three planes corresponding to the zero scattering order ($n=0$). The scattering occurs in the plane perpendicular to the chain axis, and, therefore, the planes are turned by an angle of $120^\circ$ relative to each other. The next diffraction orders ($n=\pm 1$, $\pm 2$, $\ldots $) are pairs of cones with axis coinciding with that of the chain. In the experiments, diffraction patterns are observed on a flat screen placed behind the samples. In the case of the normal incidence of light onto a sample, i.e., onto the systems of chains, zero-order scattering planes are projected on the screen as strips. If we take into account that the line of intersection between a cone and the flat screen is described by a hyperbola, we must observe in addition to the strips three pairs of arcs in the case of the first-order scattering and three more pairs of arcs for each higher order when the $|n|\lambda \le a$ relation is valid (Figs.~\ref{fig:schemes} and~\ref{fig:SEM}).

Finally, we consider a honeycomb photonic structure that demonstrates the honeycomb tiling of the plane. The honeycomb hexagonal lattice can be regarded as two interleaving triangular lattices. Because the cell boundary lies at the center of the honeycomb structure, it is necessary to take into account in the summation the shift of the zero cell for each of the parallelograms. As a result, we obtain the structural factor of honeycomb lattice that is constituted by six summands, which reflects the presence of two atoms (scatterers) per unit cell:
\begin{eqnarray}
\label{eq:SfactGraphene}
S(\mathbf{q}) & = S_{12} (\mathbf{q})\left\{e^{i\mathbf{q}\frac{\mathbf{a}_{1} -\mathbf{a}_{3} }{3} } +e^{i\mathbf{q}\frac{\mathbf{a}_{2} -\mathbf{a}_{3} }{3} } \right\} +
S_{23} (\mathbf{q})\left\{ e^{i\mathbf{q}\frac{\mathbf{a}_{2} -\mathbf{a}_{1} }{3} } + \right. \nonumber \\
& \left. + e^{i\mathbf{q}\frac{\mathbf{a}_{3} -\mathbf{a}_{1} }{3} } \right\}+S_{31} (\mathbf{q})\left\{e^{i\mathbf{q}\frac{\mathbf{a}_{3} -\mathbf{a}_{2} }{3} } +e^{i\mathbf{q}\frac{\mathbf{a}_{1} -\mathbf{a}_{2} }{3} } \right\}.
\end{eqnarray}
 On performing a number of transformations and simplifications, we can reduce formula (\ref{eq:SfactGraphene}) to
\begin{equation} \label{eq:SfGra}
\left| S(\mathbf{q})\right|^{2} \approx 4N^{2} \frac{\sin ^{2} (2N\zeta )}{\sin ^{2} (\zeta )} \cos ^{2} \frac{\zeta }{3}
\end{equation}
 for the vertical plane.

\begin{figure*}[!t]
\includegraphics{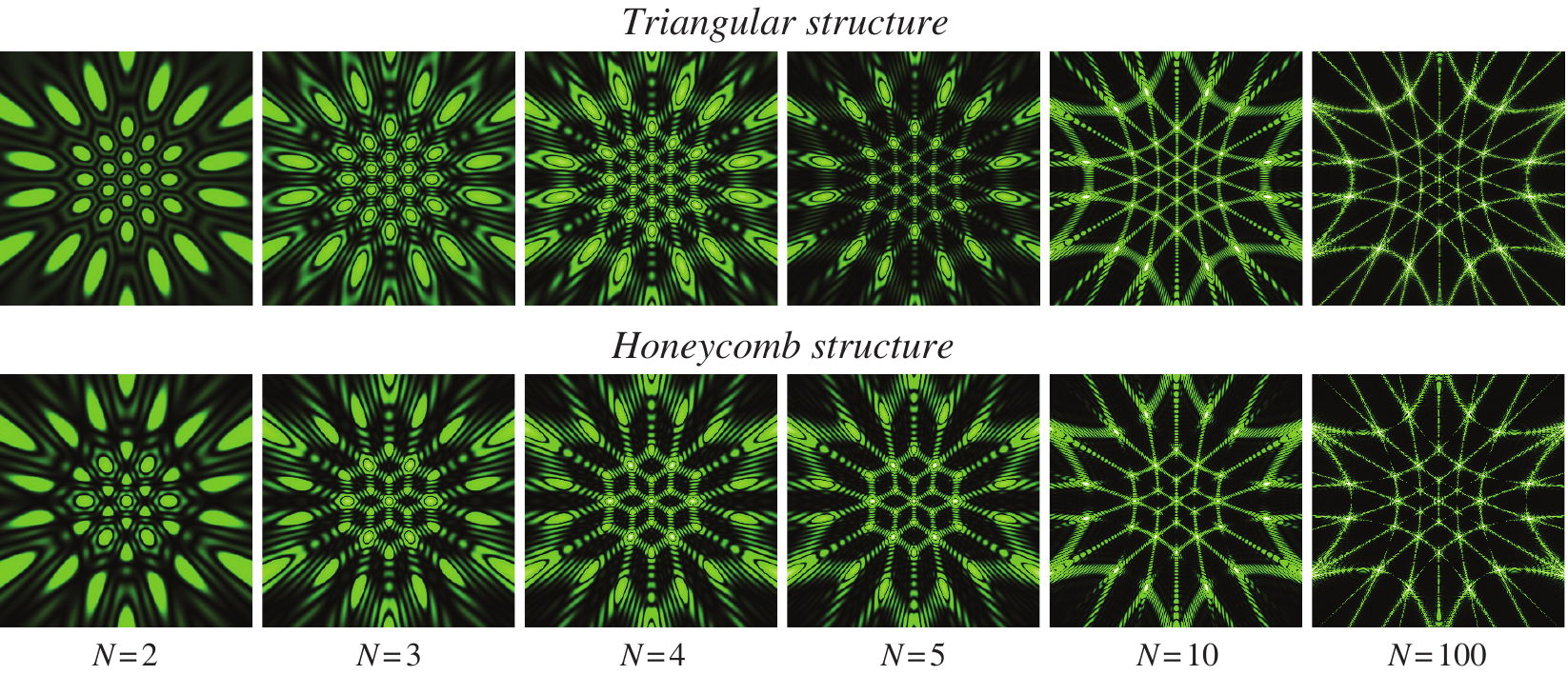}
\caption{Diffraction patterns calculated for triangular (upper row) and honeycomb (lower row) photonic structures on a flat screen positioned behind the sample. The number of scatterers $N$ is indicated above each pattern.}
\label{fig:DiffrN}
\end{figure*}

Compared with expression (\ref{eq:SfHex}) for the triangular lattice, we obtain an additional multiplier $4 \cos ^{2} \frac{\zeta }{3} $, which markedly changes the diffraction pattern. This multiplier becomes zero at $\frac{\zeta }{3} =\frac{\pi }{2} +\pi m$ and suppresses the diffraction at certain angles, both on planes and on cones. Accordingly, breaks in straight lines and hyperbolas are observed on the screen, with one of these marked by the red circle in Fig.~\ref{fig:schemes}c [see also~\ref{fig:SEM}b]. Thus, the diffraction pattern of the honeycomb lattice contains all the above features of diffraction on triangular lattice. All these specific features follow from formula (\ref{eq:SfGra}): the principal maxima [square in Fig.~\ref{fig:schemes}c] are associated with the zeros of the denominator $\sin ^{2} (\zeta )$, the additional maxima are [rectangle in Fig.~\ref{fig:schemes}c] are related to the periodic function $\sin ^{2} (2N\zeta )$ in the numerator, and, finally, the breaks characteristic of only the honeycomb lattice are associated with the zeros of the complementary function $\cos ^{2} \frac{\zeta }{3} $. These breaks in the diffraction patterns are due to the destructive interference of waves scattered from two interleaving triangular lattices forming the honeycomb lattice. The manner in which diffraction patterns are transformed with increasing number $N$ of scatterers is reflected in Fig.~\ref{fig:DiffrN} for triangular (upper row) and honeycomb (lower row) lattices.

%%%%%%%%%%%%%%%%%%%%%%%%%%%%
%% Figure 4
%
%%%%%%%%%%%%%%%%%%%%%%%%%%%%

\section{Experimental studies of optical diffraction}

%%%%%%%%%%%%%%%%%%%%%%%%%%%%
%% Figure 5
\begin{figure*}[!t]
\includegraphics[width=12cm]{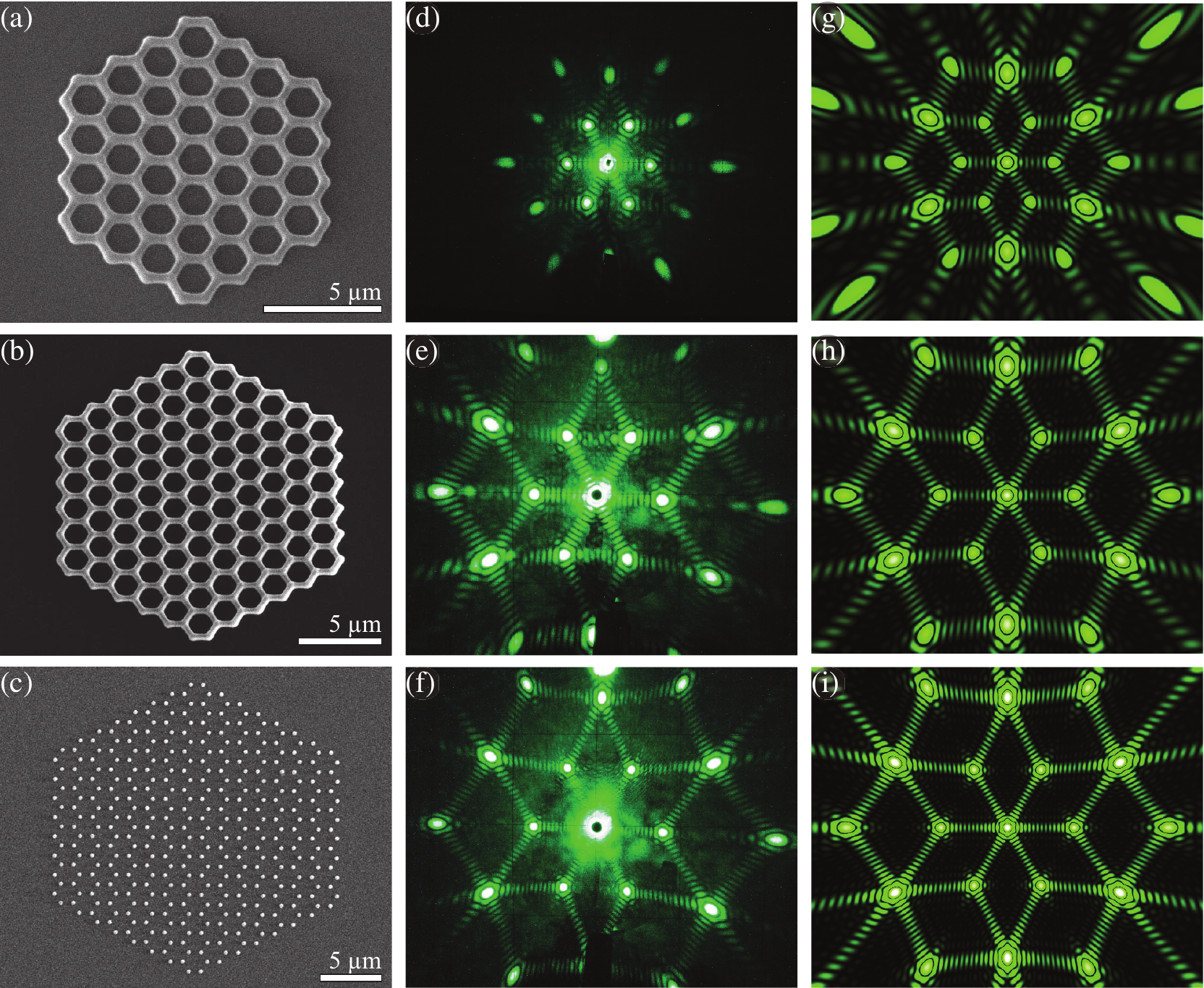}
\caption{(a-c) SEM images of the honeycomb photonic structures fabricated by the two-photon polymerization method with constant parameter $a=1\mu $m. (d-f) Experimental diffraction patterns obtained from structures (a-c), respectively. The patterns are observed on a flat screen positioned behind the sample. $\lambda =0.53\mu $m. (g-i) Diffraction patterns calculated for samples (a-c), respectively.}
\label{fig:Experiment}
\end{figure*}
%
%%%%%%%%%%%%%%%%%%%%%%%%%%%%

The diffraction patterns are examined visually and are photographed with the samples illuminated with monochromatic polarized light ($\lambda =0.53\mu $m) at the normal incidence of the laser beam onto a sample. A photograph of the installation with the screen and the diffraction pattern is shown in Fig.~\ref{fig:schemes}d. The optical system provided a full exposure of the sample, so that all the particles scattered light with the same intensity. The results of an experimental study of the diffraction of monochromatic light are presented in Figs.~\ref{fig:Experiment}d-\ref{fig:Experiment}f, together with the SEM images of the corresponding structures Figs.~\ref{fig:Experiment}a-\ref{fig:Experiment}c and calculated diffraction patterns Figs.~\ref{fig:Experiment}g-\ref{fig:Experiment}i. The two-photon polymerization method is used to fabricate a large number of samples with honeycomb structure, which differed in the number of cells, thickness of rods forming the structure, and formation principles of the structure itself. For example, Fig.~\ref{fig:Experiment}c shows a SEM image of the structure in which each honeycomb cell is formed by six voxels having planar size of about 280 nm. The number $N$ of honeycomb cells at the sample boundary is four for sample shown in Fig.~\ref{fig:Experiment}a, six for sample in Fig.~\ref{fig:Experiment}b, and eight for sample in Fig.~\ref{fig:Experiment}c. The lattice constant is the same for all the samples, $a=1\mu $m.

Surprisingly, on a screen placed just behind the sample, we observe by naked eye a fine structure of the diffraction patterns from finite-size samples, which makes it possible to characterize the shape and exactly determine the number of honeycomb cells. In accordance with the diffraction condition $|n|\lambda \le a$, the scattering of light on the samples at $a=1\mu $m must give rise to three planes (zero diffraction order, $n=0$) and three pairs of cones (first diffraction order, $n=\pm 1$). Accordingly, three strips and six arcs broken after every three principal maxima must be observed on the screen situated behind the sample. Indeed, patterns of this kind are observed experimentally, with both strips and arcs losing their intensity after every three principal maxima. This result is observed for all the three structures, with the dropout of fragments of strips and arcs most clearly demonstrated by Fig.~\ref{fig:Experiment}f.
It is noteworthy that the presence of several atoms in a unit cell may lead not only to the intensity zeroing in the fine structure, but can also suppress the strong maxima, as in the case of the face centered cubic lattice \cite{rybin2007high} in which the diffraction is observed for reflections with Miller indices of the same parity only.

It should also be noted that the fine structure of diffraction patterns associated with the function $\sin ^{2} (2N\zeta )$ in the numerator of the structural factor, is manifested in all experimental diffraction patterns. A counting of separate maxima yields a result exactly coinciding with the theoretical value: the number of additional maxima between two principal maxima is $2N-2$, which, being combined with the principal maxima themselves, gives the number $2N$. Therefore we obtain a unique chance to observe by a naked eye the fine structure of the diffraction images and to calculate the number of nano-particles and nano- honeycombs of finite structures without SEM images. When the number of honeycomb cells grows, isolated reflections start to overlap and finally merge into continuous, well-studied diffraction patterns (Fig.~\ref{fig:DiffrN}).

\section{Transmission spectra of photonic honeycomb lattice}

%%%%%%%%%%%%%%%%%%%%%%%%%%%%
%% Figure 6
\begin{figure}[!t]
\includegraphics{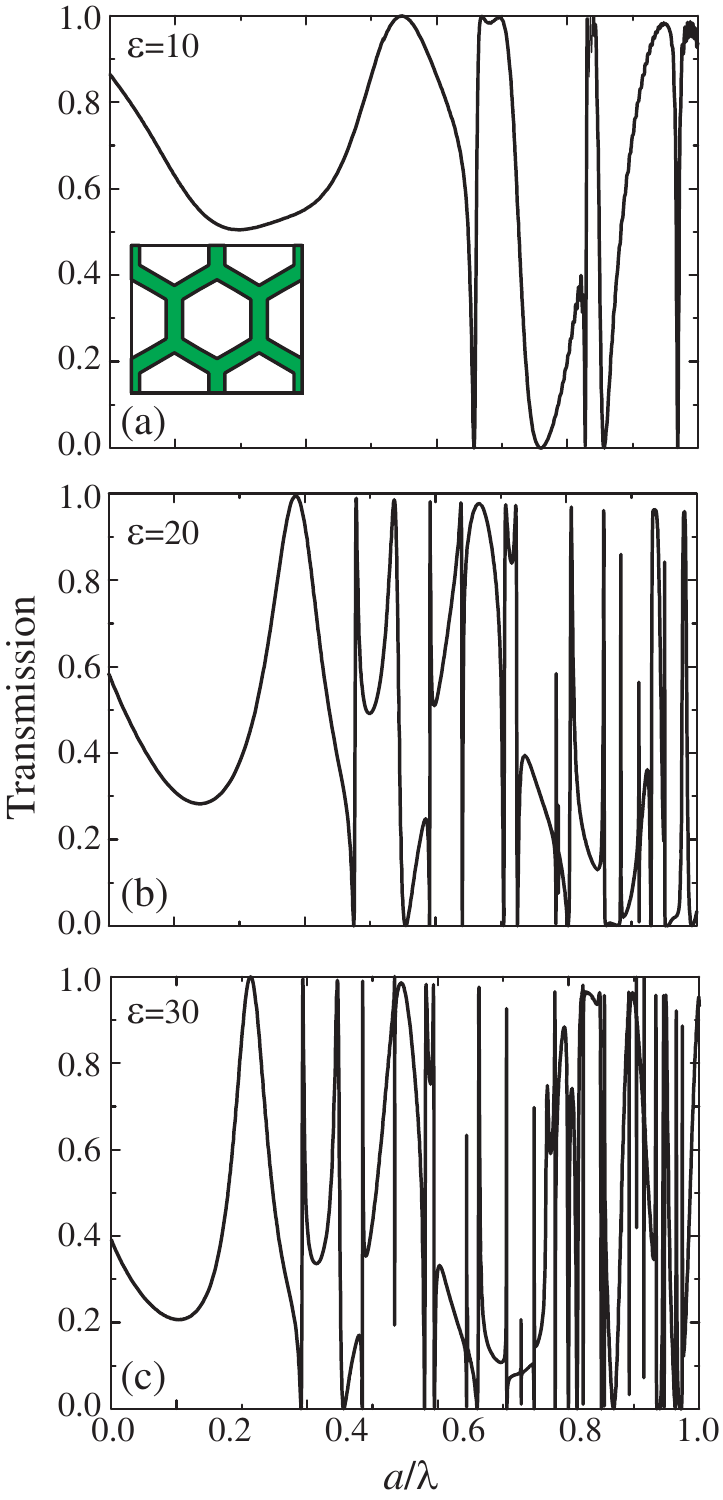}
\caption{Transmission spectra of honeycomb photonic crystal slab for different dielectric permittivity $\varepsilon=10$ (a), $\varepsilon=10$ (b), and $\varepsilon=10$ (c). The unit cell is shown in the insert of (a).}
\label{fig:Transmission}
\end{figure}
%
%%%%%%%%%%%%%%%%%%%%%%%%%%%%

We examine the transport properties of honeycomb photonic structure that is photonic analog of graphene. In this section we simulate transmission spectra of the honeycomb slab by using CST Microwave Studio software. The structure is composed from dielectric rods with a circular cross-section of radius $r=0.2a$, where $a$ is the lattice constant. The dielectric rods are connected to form the graphene structure (see insert of Fig.~\ref{fig:Transmission}a). The light is assumed to incident at the normal direction to the honeycomb slab. Transmission spectra of the sample are shown in Fig.~\ref{fig:Transmission} for several values of dielectric constant $\varepsilon=10$, 20, 30. The spectra exhibit a number of asymmetric resonances with Fano profiles superimposed on a slow varying sine background.  With the dielectric constant increasing the background period decreases and resonance lines shift to the lower frequencies.

We discuss two types of features in the transmission spectra. The first one is the slow sinusoid background being conventional Fabry-Perot oscillations on honeycomb slab boundaries. The slab thickness is $0.4a$ that is relatively small value. For higher values of dielectric constant the optical thickness increases and the background period decreases. Now we discuss the second type of feature, i.e. resonances with Fano profiles. Photonic crystal slabs are known to support in-plane guided modes~\cite{saleh2007fundamentals}, which is protected from the leakage to outside by the in-plane momentum conservation, since they lay below the light cone. However, in periodic systems any wave vector can be transformed to other wave vectors by the addition of a reciprocal lattice vector. Thus, at certain frequencies the incident wave does couple with the guided modes, which manifested in the spectra as resonant profile.

Similar spectra were observed a century ago in metallic grids and the effect is known as Wood's anomalies. Ugo Fano suggested that the resonant features in spectra are due to the excitation of surface waves \cite{fano1941theory}, that is surface plasmon polaritons interacting with the incident wave via the transformation of the incident's wave vector by the additional vector of reciprocal lattice. The asymmetric profile is a result of interference between reflected wave and the surface mode, since the phase of resonant mode changes by $\pi$ in the narrow interval of the line width, whereas the phase of reflected wave can be assumed as constant in this interval. In photonic crystal the similar spectra were analyzed by Fan and Joannopoulos \cite{fan2002analysis}. The guided resonances are strongly confined with the dielectric slab, until the light escapes to the free space modes by the wave vector transformed with the reciprocal lattice vector. However when we study transmission or reflection spectra the in-plane component of wave vector is given by the incident wave and the coupling with the guided modes is allowed for a discrete set of frequencies only.

\section{From photonic lattices to metasurfaces}

%%%%%%%%%%%%%%%%%%%%%%%%%%%%
%% Figure 7
\begin{figure*}[!t]
\includegraphics[width=16.3cm]{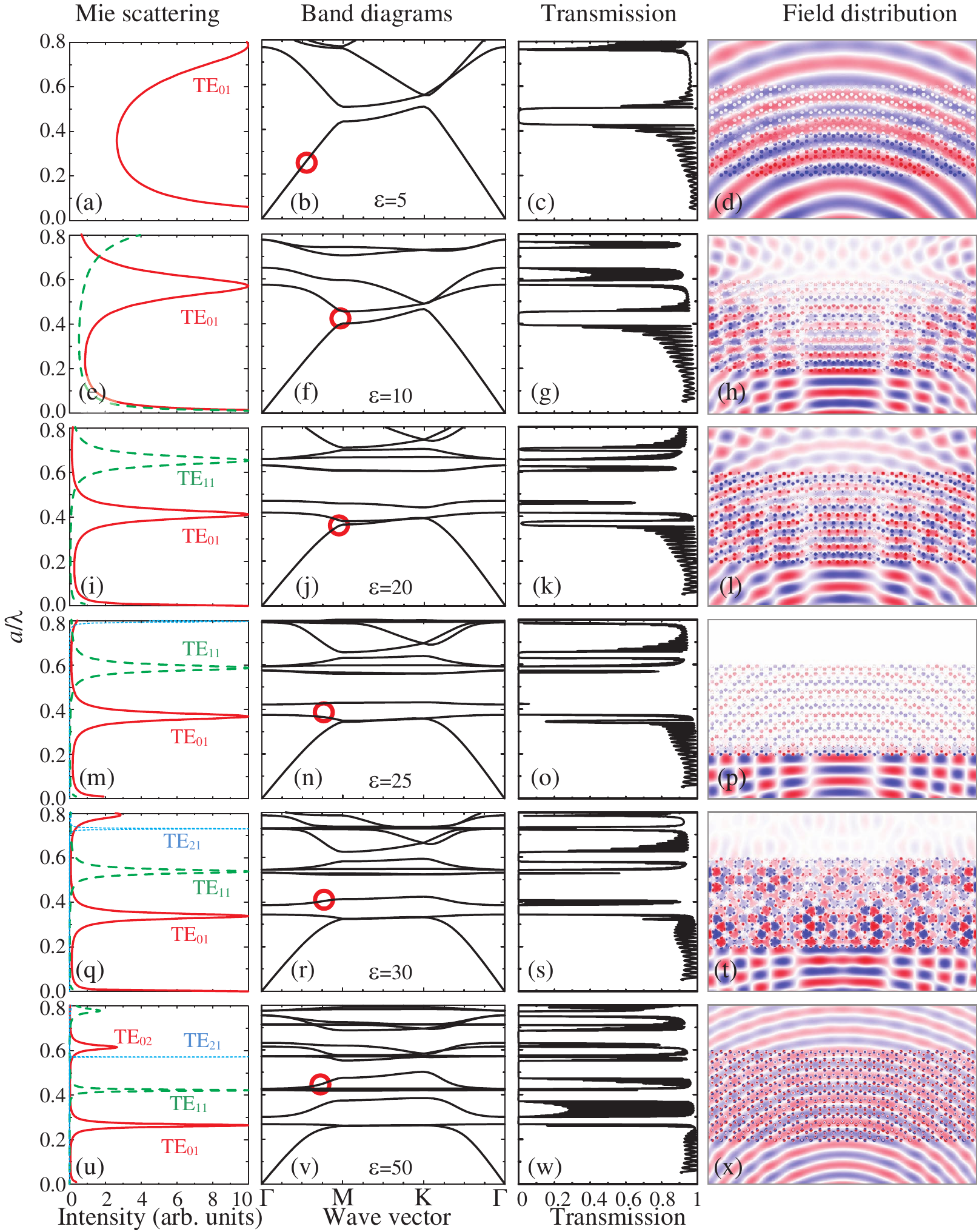}
\caption{(Left-hand column) Calculated Mie scattering efficiency $Q_{sca,n} $ for an isolated dielectric circular rod for TE${}_{nk}$ modes. (Second column) The band structure for 2D honeycomb lattice of rods with $r=0.2a$ in air ($\varepsilon _{\rm air} =1$) for the TE polarization. (Third column) The transmittance calculated for 10 lattice layers of the 2D honeycomb structure of rods in air for the TE polarization. The frequency and wave vector are plotted in dimensionless units $a/\lambda $, where $\lambda$ is vacuum wavelength and $a$ denotes the lattice constant. (Right-hand column) Results of numerical calculations for the $H_z$ component of the TE polarized electromagnetic field that has been scattered by 10 lattice layers of the 2D honeycomb structure of rods. The calculations are performed for parameters ($\left| \mathbf{k}\right|$, $a/\lambda $) emphasized by circles in the second column. (a-d) $\varepsilon = 5$, (e-h) $\varepsilon = 10$, (i-l) $\varepsilon = 20$, (m-p) $\varepsilon = 25$, (q-t) $\varepsilon = 30$, (u-x) $\varepsilon = 50$.
}
\label{fig:FullPicture}
\end{figure*}
%
%%%%%%%%%%%%%%%%%%%%%%%%%%%%

Previously we investigated a transition in light scattering regimes between photonic crystals and dielectric metamaterials analyzing the physics of Mie and Bragg resonances \cite{rybin2015phase}. The idea was based on the paper by O'Brien and Pendry \cite{o2002photonic} that demonstrates theoretically that a negative permeability $\mu <0$ can be obtained due to Mie resonance in a 2D structure with square lattice composed of dielectric circular rods with the magnetic field polarized along the axes of the rods (TE-polarization) \cite{o2002photonic}. We demonstrated theoretically and experimentally that a 2D periodic photonic structure with square lattice transforms into a metamaterial when the TE${}_{01}$ Mie gap opens up below the lowest Bragg bandgap where the homogenization approach can be justified and the effective permeability becomes negative \cite{rybin2015phase}.

Here we expand this concept on the 2D photonic structure with honeycomb lattice of parallel dielectric circular rods infinitely long in the $z$ direction with the radius $r$ and real frequency independent permittivity $\varepsilon $. The structures with the lattice constant $a$ can be characterized by the filling ratio $r/a$. We have calculated four sets of spectroscopic data depending on $\varepsilon $ (Fig.~\ref{fig:FullPicture}), namely: (i) The spectra of the Mie scattering $Q_{sca}$ by an isolated rod. For TE polarization, the far-field scattering can be described by circular Lorenz-Mie resonant coefficients $a_{n}$ corresponding to magnetic moments \cite{g120}. We calculated numerically spectra of the Mie scattering efficiency $Q_{sca,n} =\frac{2}{x} \left|a_{n} \right|^{2} $ for the dipole TE${}_{0k}$ and higher multipole TE${}_{nk}$ ($n\ge 1$, $k\ge 1$) modes in the range of normalized frequencies $a/\lambda$ from 0 to 0.8. (ii) The photonic band structure of an infinite 2D lattice composed of rods along the $\Gamma\to$M$\to$K$\to\Gamma$ direction of the Brillouin zone. The dispersion relation of the eigenmodes we calculated using the plane wave expansion method \cite{g405}. The photonic band structures were obtained with 128 by 128 plane waves. (iii) The transmission spectra of a 2D lattice of 10 honeycomb layers in length were calculated by using the CST Microwave Studio software for the wave vector of the incident beam parallel to the $\Gamma\to$M direction. (iv) The structure of the magnetic field ($H_{z}$ component) inside and around a 10 layers honeycomb structure excited by a point source with the TE polarization.

All data sets have been calculated for a filling ratio $r/a=0.2$ and a wide range of the rod permittivity $1\le \varepsilon \le 50$ with the step of $\Delta \varepsilon =1$. Note that at $r/a=1/(2\sqrt{3}) \approx 0.289$ all rods touched each other and at $r{\rm /}a>1/(2\sqrt{3}) $ the rods penetrate into each other and the structure appears as inverted one of the air holes in the dielectric matrix. The obtained data allows us to analyze precisely the evolution of the low-frequency region of the photonic band structure formed by a complicated mixture of Mie and Bragg resonances and propagating modes. For low-contrast honeycomb structure, all Mie resonances are located is a frequency range higher than the lower Bragg gap [Fig.~\ref{fig:FullPicture}a-\ref{fig:FullPicture}h)]. The calculations reveal strong decreasing of the Mie frequencies and narrowing of the resonant Mie bands with $\varepsilon$ increasing. The Mie resonances of single rods, correspondingly Mie bands of the photonic structure and the transmission Mie dips demonstrate shift from the higher to low frequencies, crossing the strongly dispersive Bragg bands. At dielectric permittivity about $\varepsilon=25$, the lowest TE${}_{01}$ Mie gap completely splits from complicated coupled Mie-Bragg band becoming the lowest gap in the spectrum [Fig.~\ref{fig:FullPicture}m-\ref{fig:FullPicture}o)]. As a result, the lowest Bragg band and the TE${}_{01}$ Mie band change their positions in the energy scale signalizing the transition from photonic crystal phase into metamaterial phase with negative permeability $\mu <0$ \cite{o2002photonic}. Since all the Bragg gaps are located at higher frequencies, any diffraction losses in the vicinity of the Mie resonance are absent.

The right-hand column in Fig.~\ref{fig:FullPicture} presents the magnetic field distribution ($H_z$ component) for a 10 layers honeycomb structure. The strong diffraction of waves at Bragg frequencies is clearly seen inside and around the structure [Fig.~\ref{fig:FullPicture}h and~\ref{fig:FullPicture}l)]. In contrast, within the TE${}_{01}$ Mie gap the transmittance is nearly zero and the structure appears as perfectly homogenous without any traces of Bragg scattering waves. It is clear evidence that the structure can be considered as homogeneous medium that is the necessary condition for appearance of the metamaterial phase \cite{andryieuski2010homogenization}. Right-hand column in Fig.~\ref{fig:FullPicture} demonstrates different scattering regimes including homogeneous transmission of waves near the second Mie band [Fig.~\ref{fig:FullPicture}x].

\section{Dirac lensing effect}

%% Figure 8
\begin{figure*}[!t]
\includegraphics[width=16.3cm]{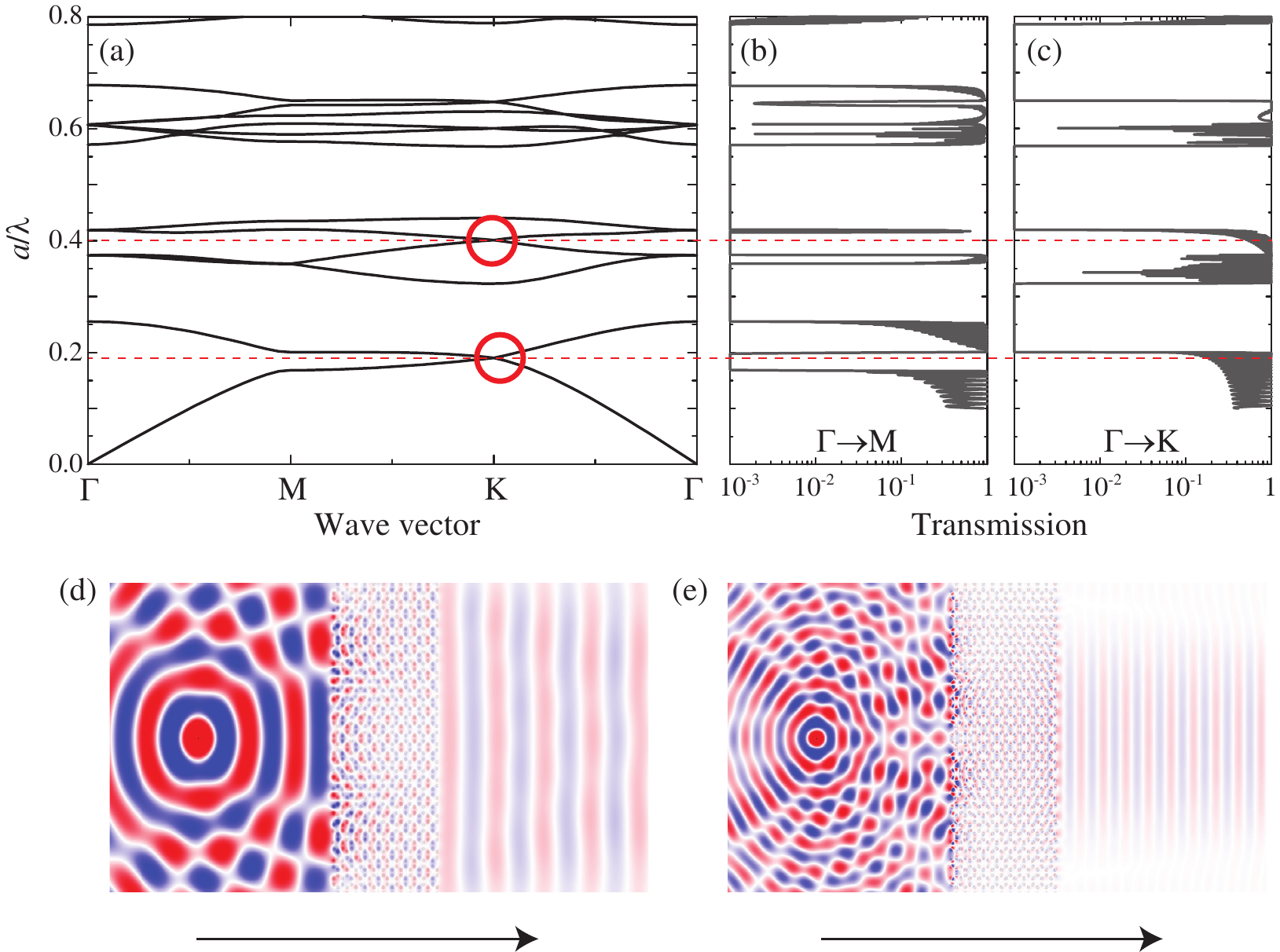}
\caption{Dirac point effects in photonic graphene. (a) The band structure of photonic graphene composed of rods with $r=0.2a$ and $\varepsilon=0.25$ in air for the TM-polarization. Red circles mark Dirac points. Transmission spectrum of photonic graphene layer with zigzag surface (b) and armchair surface termination (c). The Dirac frequencies are shown by red dashed line across panels (a)-(c). Dirac lensing effect at the Dirac point at $a/\lambda=0.19$ (d) and $a/\lambda=0.40$ (e). The point source is placed at the distance of $15a$ from the left boundary of the layer. The direction of the beam which wavefronts are parallel planes is shown by arrows in the bottom of panels (d) and (e).}
\label{fig:Dirac}
\end{figure*}
%
%%%%%%%%%%%%%%%%%%%%%%%%%%%%

The symmetry analysis of the honeycomb structure predicts the existence of doubly degenerate states at the $K$ point in reciprocal space. The triply degenerate states in the empty lattice approximations split into one of $A$-type and two degenerate $E$-type states \cite{sakoda2004book}. Since the $K$-point does not have inversion symmetry the linearly dispersion conditions occurs and two bands touch at the so-called Dirac point that is an apex of the Dirac cones.

In the case of photonic structures, Dirac cones are of interest because of the existence of robust surface states due to the breaking of parity and time-reversal symmetry \cite{raghu2008analogs}. Also in the vicinity of the Dirac point hexagonal structures exhibit intrigue transport properties \cite{sepkhanov2007extremal,diem2010transmission}. 
In particular, it was found that the transmission $T$ has an extremum at the frequency of the mode degeneration in the $K$ point where $T$ demonstrates a pseudodiffucive $1/L$ dependence on the longitudinal dimension $L$ of the sample. We notice that nontrivial transport properties was found on related $K$ point of three-dimensional face centered cubic structures \cite{moroz2011multiple}.

Figure \ref{fig:FullPicture} shows that the photonic band diagrams for TE polarization indeed exhibit band degenerations on the $K$ point, however all these intersections of the dispersion branches take place at the allowed frequencies for other directions. Thus those intersections are unsuitable for observation of strong effects on the transport properties.

Figure~\ref{fig:Dirac}a demonstrates the photonic band diagram calculated for TM-polarized wave traveling in the structure composed of circular rods with  parameters $r=0.2a$ and $\varepsilon=25$. Two Dirac cones are clearly seen in the Band diagram at the frequencies $a/\lambda=0.19$ and  $a/\lambda=0.40$. By using the CST Microwave Studio we calculate transmission spectra for both $\Gamma\to M$ (the plane wave incidents normally to the zigzag edge) and $\Gamma\to K$ directions (the plane wave incidents normally to the armchair edge).
Transmission spectrum for the $\Gamma\to M$ direction demonstrates stop-bands in the frequency intervals $0.17<a/\lambda<0.20$, $0.37<a/\lambda<0.41$ and $0.42<a/\lambda<0.57$ and other higher frequency gaps (Fig.~\ref{fig:Dirac}b). The transmission for  $\Gamma\to K$ direction exhibits stop-bands in the intervals $0.20<a/\lambda<0.32$ and $0.42<a/\lambda<0.57$ (Fig.~\ref{fig:Dirac}c). We notice that the high frequency edge of the stop-band in $\Gamma\to M$ direction matches the lower frequency stop-band edge in $\Gamma\to K$ at $a/\lambda=0.20$. Therefore, the waves with the frequencies of the stop-band for $\Gamma\to M$ directions are allowed for the propagation in the $\Gamma\to K$ directions. Note that at the Dirac point frequency $a/\lambda=0.19$ the only allowed direction is $\Gamma\to K$ (red dashed line in Fig.~\ref{fig:Dirac}a). The similar situation occurs for the second Dirac point at $a/\lambda=0.40$.

We study how photonic graphene layer interacts with the TM-polarized wave from a point source with the Dirac point frequencies $a/\lambda=0.19$ and $a/\lambda=0.40$. The wave incidenting on the photonic graphene layer surface can enter the structure only in the normal ($\Gamma\to K$ for armchair orientation) direction. As a result the passing wave at the opposite side of the layer forms a beam with flat phase surfaces, which propagates in the normal direction. The simulated electric field patterns are shown in Fig.~\ref{fig:Dirac}d for the first Dirac point and in Fig.~\ref{fig:Dirac}e for the second Dirac point. Therefore, photonic graphene acts as a lens focusing a circular wave into a beam, which we call \emph{a Dirac lensing} effect. The Dirac lens represents an example of novel photonic device for manipulations with light having only several wavelength in thickness.

We notice here that the Dirac cone feature is not related to lattices with C$_6$ symmetry only, where it is protected by the symmetry. By employing of accidental degeneracy, a square lattice of dielectric cylindrical rods was shown \cite{huang2011dirac} to exhibit Dirac cone dispersion at the center of the Brillouin zone (the $\Gamma$ point). In the case of $|\mathbf{k}|\ll 1$ an effective medium description is possible \cite{huang2011dirac} and it predicts near zero effective parameters ($\varepsilon\approx 0$ and  $\mu\approx 0$) associated with tunneling through distorted channels, cloaking, lensing and other effects \cite{liberal2017near}. Albeit the Dirac cones in $K$ points cannot be described by an effective medium theory, the Dirac lensing effect considered above is robust being protected by the honeycomb lattice symmetry in contrast to the Dirac point in the $\Gamma$ point.

\section{Conclusions}

We have demonstrated the results of comprehensive studies of electromagnetic properties of honeycomb structures. We employ two-photon polymerization technique \cite{kawata2001finer,farsari2009materials} to fabricate various samples of honeycomb structures and study both theoretically and experimentally optical Laue diffraction of monochromatic light analyzing their dependence on the number of honeycomb cells in the sample. By choosing the lattice parameters and laser wavelength, we visualize diffraction patterns on a flat screen placed behind the sample. A detailed interpretation of the complex diffraction patterns allow the step-by-step analysis of the structural factor $\left|S(\mathbf{q})\right|^{2} $ in the Born approximation, first considering one-dimensional arrays of scatterers, then employing two-dimensional structure with the symmetry of graphene formed by honeycombs with two scatterers per unit cell. A surprisingly strong optical diffraction was observed from microscopic samples formed by nanoscale elements, which allow a detailed analysis of the patterns and geometry of the samples. We have identified three types of specific features in two-dimensional diffraction patterns. Two kinds of these, the principal and additional maxima, are observed for the structures with one scatterer per unit cell, and the third one, the dropout of diffraction reflections and breakup of strips and arcs after every three principal maxima is observed only in the diffraction patterns from graphene lattices. We obtained an excellent agreement between numerical calculations and experimental data.

In the theoretical studies of the transmission of both honeycomb structures and 2D honeycomb lattice of parallel dielectric circular rods we demonstrated an important role of the Mie resonanses. The calculations reveal a strong decrease of the Mie eigenfrequencies and narrowing of the resonant Mie bands with $\varepsilon $ increasing. The Mie resonances of individual rods formed non-dispersive flat bands in the photonic structure and Mie photonic gaps in the energy spectrum of the 2D honeycomb photonic structure. The structure transforms into a metamaterial when the lowest TE${}_{01}$ Mie gap opens up below the lowest Bragg bandgap. This transformation leads to the homogenization of the periodic honeycomb structure. Within the TE${}_{01}$ Mie gap, the effective magnetic permeability is negative ($\mu <0$) and all waves in the medium are evanescent.

For 2D photonic graphene structure we have considered two Dirac points at the $K$ point of the 2D Brillouin zone for the TM polarization. At the Dirac point frequencies the only allowed direction of waves propagation is $\Gamma\to K$. We have demonstrated the effect of Dirac lensing: at the Dirac frequency for the TM polarization a layer of 2D photonic graphene in the armchair ($\Gamma\to K$) scattering geometry converts a wave from point source into a beam with flat phase surfaces. This will introduce a concept for manipulating light by nanoscale photonic devices with Dirac features.

\section*{Acknowledgments}

We acknowledge support by the Russian Science Foundation (Grant 15-12-00040, fabrication of optical honeycomb samples by two-photon polymerization, theoretical and experimental studies of optical diffraction) and a support by the Ministry of Education and Science of Russian Federation (Project 3.1500.2017/4.6) and the Russian Foundation for Basic Research (grant 15-02-07529).

%\bibliographystyle{revtex}
%\bibliography{graphene}

\end{document}